# THE EFFECT OF OIL PRICE ON UNITED ARAB EMIRATES GOODS TRADE DEFICIT WITH THE UNITED STATES

Osama D. Sweidan & Bashar H. Malkawi

**ABSTRACT**

We seek to investigate the effect of oil price on UAE goods trade deficit with the U.S. The current increase in the price of oil and the absence of significant studies in the UAE economy are the main motives behind the current study. Our paper focuses on a small portion of UAE trade, which is 11% of the UAE foreign trade, however, it is a significant part since the U.S. is a major trade partner with the UAE. The current paper concludes that oil price has a significant positive influence on real imports. At the same time, oil price does not have a significant effect on real exports. As a result, any increase in the price of oil increases goods trade deficit of the UAE economy. The policy implication of the current paper is that the revenue of oil sales is not used to encourage UAE real exports.

*Corresponding Author:* **Osama D. Sweidan** *Dept of Finance & Economics, College of Business Administrations, PO Box 27272, University of Sharjah, Sharjah UAE*
*Tel: 971 50 9832770, Email: osweidan@sharjah.ac.ae*
**Bashar H. Malkawi,** *Dept of Private Law, College of Law, PO Box 27272, University of Sharjah Sharjah UAE, Email: bmalkawi@sharjah.ac.ae*

## 1. INTRODUCTION

Since the Arab oil embargo (1973–1974), exploring the relationship between oil price fluctuations and macroeconomic performance attracted the attention of large number of economists. Technically, economists focused on testing to what extent oil price fluctuations are responsible for establishing recessions; and on policy reaction of how economies should adapt with the new situation. Those two questions appear again significantly because of the latest massive increase of oil price of 2008 and now in 2011.

Earlier studies (e.g. Berndt and Khaled, 1979, Berndt and Wood, 1979 and Wilcox, 1983,) which followed the oil price shock of (1973-1974) made a link between price of energy and capital accumulation. They estimated aggregate production functions with models allowing for varying degrees of substitutability among factors of production. They found a net complementary relationship between oil price and capital accumulation in the industrialized countries. Hence, increase in oil price decreases investment growth rate. During the period followed the oil price shock of (1973-1974), the basic understanding of oil price is its significant impact and causality relationship with the macroeconomic variables. Sachs (1981) demonstrated that higher oil prices in (1973-74) led to massive OPEC surplus matched by large deficits in developed and developing countries. Consequently, income and consumption are affected in these groups of countries. At that time, economists believed implicitly that oil price has a significant impact and causality relationship with macroeconomic variables. This point of view started to alter with a new generations of studies (e.g., Rasche and Tatom, 1977; Rasche and Tatom, 1981; Darby, 1982; Hamilton, 1983; Burbidge and Hassison, 1984; Gisser and Goodwin, 1986) who succeeded to establish another link that an unanticipated positive oil price shock has a negative effect on output. However, they failed to prove a direct causality relationship between price of oil movements, international recessions, and U.S stagflation.

Barsky and Kilian (2002) highlighted the same conclusions. They stated that volatility of oil price is likely to matter less for U.S. macroeconomic performance than has commonly been thought. They illustrated that oil price shocks may contribute to recessions and inflation rate without necessarily being fundamental. They presented several large spikes in inflation rate and recessions which are clearly unrelated to oil events. Moreover, Hunt (2006) proved an increase in the price of oil is able to generate inflation persistence similar to that seen in 1970s. On the contrary, he found that energy price shocks cannot create the type of stagflation observed in 1970s. Korhonen and Juurikkala (2007) investigated the determinants of equilibrium real exchange rates in a sample of oil-dependent countries over the period (1975-2005). They concluded that oil price has a noticeable statistically significant effect on real exchange rates. Technically, higher oil price leads to real exchange rate appreciation. On the other hand, real per capita gross domestic product does not have an obvious effect on real exchange rate. Husain et al. (2008) assessed the impact of oil price shocks on the non-oil economic cycle in oil-exporting countries. They discovered that oil price changes have a significant impact on the economic cycle but only through their impact on fiscal policy. Lorde et al. (2009) empirically inspected the macroeconomic effects of oil price fluctuations on small open oil-producing countries, i.e. Trinidad and Tobago. They concluded that oil price is a major determinant of economic activity and unanticipated shocks to oil price and oil price volatility generate significant swings in the economy. Galesi and Lombardi (2009) explored to what extent oil and food price shocks pass on to the inflationary stance and real economy. They concluded that the direct inflationary impacts of oil price shocks influence mostly developed countries while less sizeable effects are detected for emerging economies. Further, food price shocks have significant inflationary direct effects, but especially for emerging economies.

   Bollino (2007) stated that the sharp increase in price of oil since 2003 has worsened the U.S. trade deficit because the volume of the U.S petroleum imports has remained constant; from August 2004 to July 2006 in spite of this increase. As a result, the value of the US petroleum deficit account increases by around 12 billion dollars. Moreover, Rebucci and Spatafora (2006) justified how an advanced oil-importing economy such as the U.S. economy adjusts to a permanent increase in oil price. As a preliminary result, they found that the rise in the price of oil increases the overall trade deficit. However, in the long run the change in the relative prices will ultimately eliminate the U.S. trade deficit. They explained this result as follows: As price of oil rises, oil imports become more expensive. Thus, households and investors have fewer resources to spend on goods and services, which lead to a deterioration in domestic demand for tradable goods. Consequently, it means a decline in the term of trade; the relative price of domestic tradable goods in terms of foreign tradable goods. The decline in the domestic demand for tradable goods can be pushed-up again by a higher foreign demand coming from an oil-exporting country through its higher oil revenues. This mechanism might help to eliminate the overall trade deficit. Cooper (2008) demonstrated that the U.S trade deficit is related directly to the large trade surpluses of oil exporting countries because of the sharp rise in oil prices since 2003. As a result, he stated that the U.S current account deficit would have been reduced significantly if the prices of oil were to return to the $24 a barrel that prevailed in 2002. Zaouali (2007) examined the impact of oil prices on the Chinese economy. She found that increasing oil prices have modest effects on the current account and gross domestic product. She explained this result by the ability of the Chinese economy to attract investment and foreign capital. Ozlale and Pekkurnaz (2010) explored the influence of oil price shocks on the current account balance for the Turkish economy. They proved that oil price shocks have a statistically significant negative effect on current account balance.

   With new generations of asymmetric research, Huang et al. (2005) examined the influence of oil price movements and its volatility on economic activities; changes in industrial production and real stock returns. They found a significant threshold effect. As a result, oil price fluctuations or its volatility have a limited effect on the economic activity if the change is below the threshold levels. On the contrary, if the fluctuations are above the threshold, then they have fundamental impact on macroeconomic variables even better than real interest rate. Mehrara (2008) investigated asymmetric relationship between oil revenues and output growth for 13 oil-exporting countries. He concluded that output responds in an asymmetric or non-linear manner for oil revenues shocks. Particularly, he discovered that the response of output growth for a negative oil price shock is larger than a positive shock. He stated that an adverse shock usually worsens the government budget.

On the contrary, a positive shock increases revenues and the quality of spending deteriorates which affects efficiency negatively. It is apparent from the above-mentioned literature review that oil price shocks attracted enormous attention after 1974. The goal of a large empirical works of many economists is to investigate the various transmission channels of oil price shock on economic activity.

The United Arab Emirates (UAE) economy relies heavily on the revenues generated from exporting oil. The relative importance of crude oil to gross domestic product during the period (1993-2008) is around 32%. This ratio climbed to around 37% in 2008 because of the considerable increase in oil price. Thus, the purpose of the current paper is similar to that goal which is to understand the effect of oil price on the United Arab Emirates goods trade deficit with the U.S. economy. The contribution of the current paper is to explore the impact of oil price shocks in a developing-oil-exporting country. We claim, based on our knowledge, that this is the first study that investigates this issue in the UAE. Moreover, the UAE lacks of such a significant research due to the fact that the UAE economy lacks long time series data. Thus, we rely on the data disseminated by the U.S. Census Bureau regarding trade in goods between the U.S. and the UAE. Although our paper focuses on a small portion of UAE trade[1], nevertheless, we recognize that the U.S. is a major trade partner with the UAE. The rest of the paper is organized as follows: Section 2 presents the UAE trade policy. Section 3 introduces the data and methodology of the current study. Section 4 presents the empirical results. Conclusions are presented in section 5.

## 2. THE UAE TRADE POLICY

The UAE became an original member of the World Trade Organization (WTO) on April 10, 1996, it had been a contracting party to the General Agreement on Tariff and Trade (GATT) since March 8, 1994[2]. The UAE main economic policy goal is to adopt sound economic policies that guarantee sustained economic growth, diversify the economy away from oil industry, create more employment opportunities for all its citizens and attract local and foreign investment. In addition, economic policymakers are working to promote a progressive economic agenda, build around economic liberalization and diversification, and promotion of the private sector role in the economy. Diversification includes developing further the emirate's tourism, media, shipping, financial, and commercial services, as well as expanding its industrial base.

Through its membership in the WTO, the UAE has full commitment and responsibility toward a more open international trade environment. For example, the UAE adopted a 5 percent average tariff for all goods. On the other hand, the maximum tariff that can be charged under international obligations stands at 15 percent. Simultaneously, the UAE applied an average tariff of 6.5 percent for agricultural goods compared to an average of 25 percent as a maximum tariff can be charged under international obligations. Further, the UAE has signed several regional, bilateral and preferential trade agreements with the goal of further trade

---

[1] Technically, on average, our study focuses on 11% of the UAE foreign trade.
[2] More details about members and observers see the following link:
   http://www.wto.org/english/thewto_e/whatis_e/tif_e/org6_e.htm.

liberalization.³ The UAE exports and imports of goods including oil and non-oil products reflect the country's broad integration and openness into the world economy. The trade policy review of UAE in 2010 shows that the exports of goods including oil and non-oil products increased approximately by 107.8 percent from 1995 to 2008, compared to 136.6 percent for imports of goods.⁴ The major trading partners of the UAE are: India, China, the U.S., Japan, Germany and the GCC countries.

## 2.1 The Structure of Good Trade between the UAE and the U.S.

The U.S Census Bureau provides detailed information regarding the trade of goods between the U.S. and the UAE. The data is classified based on the North American Industry Classification System (NAICS). We focus the analysis of this section over the period (2000-2009) and we utilize annul data classified based on 3-digit NAICS.

The data illustrates that the value of the UAE export of good to the U.S. is around $12.4 billion during the period (2000-2009). During the same period, the value of UAE import of goods from the U.S. is roughly $24.4 billion. Table (1) presents some details about the goods traded between the UAE and the USA during the above-mentioned period.

**Table 1: Trade of goods between the UAE and the U.S. over the period (2000-2009), 3-digit NAICS**

| | | |
|---|---|---|
| Agricultural Products | Leather and Allied Products | Computer and Electronic Products |
| Livestock and Livestock Products | Wood Products | Electrical Equipment, Appliances, and Component |
| Forestry Products | Paper | Transportation Equipment |
| Fish, Fresh, Chilled, or Frozen and Other Marine Products | Printed Matter and Related Products | Furniture and Fixtures |
| Oil And Gas | Petroleum and Coal Products | Miscellaneous Manufactured Commodities |
| Minerals and Ores | Chemicals | Newspapers, Books and Other Published Matter |
| Food and Kindred Products | Plastics and Rubber Products | Waste and Scrap |
| Beverages and Tobacco Products | Nonmetallic Mineral Products | Used or Second-hand Merchandise |
| Textiles and Fabrics | Primary Metal Manufacturing | Goods Returned to Canada and the U.S. |
| Textile Mill Products | Fabricated Metal Products | Special Classification Provisions |
| Apparel and Accessories | Machinery, Except Electrical | |

Source: The U.S. Census Bureau.

---

³ For more details see Trade Policy Review of UAE for 2010, pages 19-24, Moore (2009) and Yerkey (2004).

⁴ The trade policy review of WTO was the result of the Uruguay Round. The objectives of the trade policy review include facilitating the smooth functioning of the multilateral trading system by enhancing the transparency of WTO members' trade policies. The frequency of the trade policy review depends on the members' share of world trade, for more details see Mah (1997).

Consequently, Table (2) introduces the average relative importance of the exports and imports between the two countries. UAE goods exports to the U.S. concentrate relatively on raw materials. On the contrary, UAE imports from U.S. focus on manufactured goods. Monthly data reveals an obvious fluctuation in the relative importance of most goods traded between the two nations. Relatively, UAE exports of goods to the U.S. fluctuate more than the UAE imports of goods from the U.S.

**Table 2: The relative importance of the major traded goods between the two nations over the period (2000-2009)**

| UAE Exports | | UAE Imports | |
| --- | --- | --- | --- |
| Item | Average Relative Importance (%) | Item | Average Relative Importance |
| Oil, Gas, Petroleum and Coal Products | 20.6 | Chemicals | 5.6 |
| Miscellaneous Manufactured Commodities | 6.9 | Computer and Electronic Products | 10.6 |
| Primary Metal Manufacturing | 17.2 | Machinery, Except Electrical | 18.6 |
| Apparel and Accessories | 18.0 | Transportation Equipment | 34.8 |
| Others | 37.3 | Others | 30.4 |
| **Total** | **100** | | **100** |

Source: The U.S. Census Bureau.

## 3. DATA AND METHODOLOGY

### 3.1 The Data

The current study aims at exploring the relationship between oil price and good trade balance between the UAE and the U.S. We believe that examining this relationship in the UAE is fundamental and necessary because the UAE is an oil exporting country and price of oil is the main source of its income. Furthermore, the UAE economy lacks such studies. The economic data base of the UAE is limited in terms of frequency and availability of time series. As a result, exploring the relationship among macroeconomic variables is limited and not widely spread in the UAE economy. Thus, we rely on external sources to acquire data. The required series of UAE exports and imports are obtained from the U.S. Census Bureau.[5] This data includes only the trade of goods between the U.S. and the UAE. Although, the analysis of the current paper focuses on around 11% of the UAE foreign trade, we believe it provides us with a significant proxy of how crude oil price affects UAE international trade and economy. In addition, we obtained export and import price indices from Bureau of Labor Statistics.[6] For the purpose of calculating real exports of the UAE, we divided the UAE nominal exports by the U.S. import index of non-manufactured goods from other "not developed" countries. Also, we computed real imports by dividing the UAE nominal imports by the U.S. export index for all commodities. Moreover, price of crude oil is obtained online.[7] The crude oil price is a simple average of three spot prices: Dated Brent, West Texas Intermediate, and the Dubai Fateh.

---

[5] The U.S. Census Bureau, available on line at: http://www.census.gov.
[6] The Bureau of Labor Statistics, available at: http://www.bls.gov/home.htm.

Figures (1) and (2), below, present a *normalized* relationship between oil price and UAE real exports and real imports, respectively, over the period (1992:09-2007:12). The main conclusion of the two figures is that the UAE real imports have a stronger consistent dynamic relationship with the crude oil price than the UAE real exports. This feature is clear in Figure (1) which shows an unambiguous divergence between oil price and real exports starting from the year 1997 and started to be more obvious from 2003. Additionally, it is apparent that UAE real exports tend to fluctuate more than UAE real imports.

**Figure 1: The UAE real exports and crude oil price**

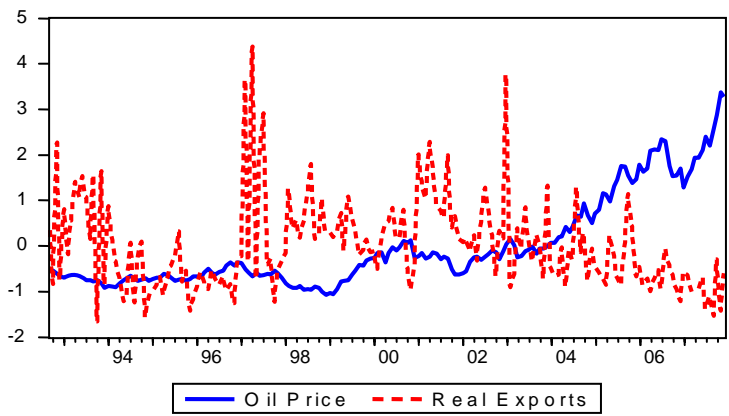

**Figure 2: The UAE real imports and crude oil price**

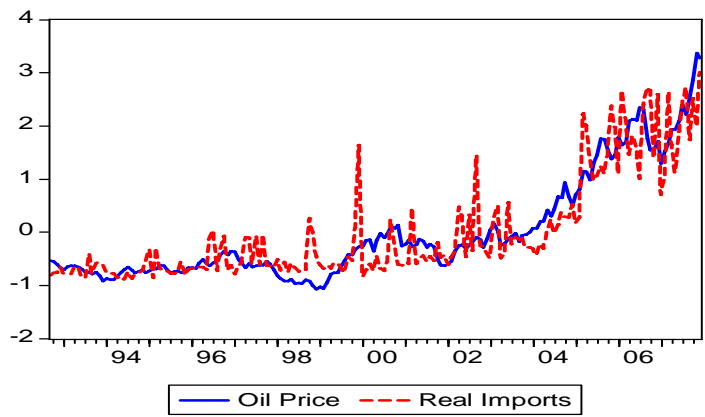

### 3.2 The Methodology

The current study uses monthly data during the period (1992:09-2007:12). The current study did not extend the data sample beyond 2007 in order to control the affects of the abrupt increase in price of oil and the consequences of the great recession on the UAE goods trade deficit. We believe our empirical options are very limited due to the lack of data. Therefore, we do not have a space to maneuver and we work with on linear models. As a preliminary step in our analysis, we explore whether the data on the level has a unit root or not. This is an essential step to avoid producing a spurious regression. Figure (3) presents a plot of each series, besides it illustrates that real imports, goods trade deficit and crude oil price have structural breaks over the period of the study. We believe that this type of structural break allows for a change in the slope or the rate of growth. Hence, we need to check it up. Dates of the structure breaks are as follow: for real imports and goods trade deficit is March 2005 and for crude oil price is October 2003. This implies that oil price structural break precedes real imports and goods trade deficit structural breaks by fifteen months. Since standard unit root tests have a reduced power if they are applied to a time series with a structural break, we utilize the unit root test by Perron (1989) that considers structural breaks. Perron (1989) model permits for a break in the slope or the rate of growth as follows:

$$\Delta Y_t = \alpha_0 + \alpha_1 SD_t + \alpha_2 t + \alpha_3 Y_{t-1} + \sum_{i=1}^{j} \gamma_i \Delta Y_{t-i} + e_t \qquad (1)$$

where $\Delta$ is the first difference, $Y_t$ stands for the time series being tested, $SD_t$ is the slope dummy, it introduces a change in the slope of change in the growth rate; $SD_t = t$ if $t >$ the structural break and zero otherwise. $t$ denotes the time trend variable, $e_t$ stands for the error term and it is normally distributed $\sim (0, \sigma^2)$.

---

[7] Index Mundi, available on line at: http://www.indexmundi.com.

**Figure 3: The variables of the model[8]**

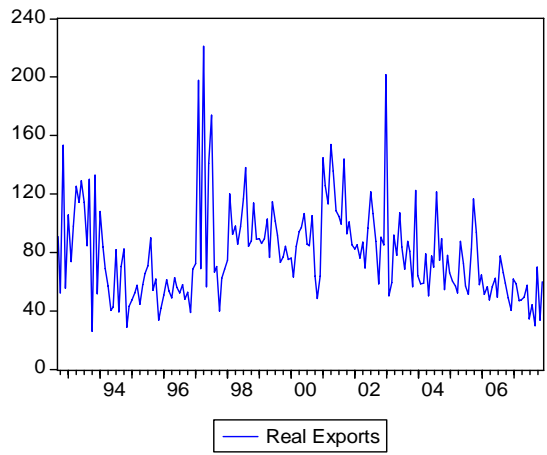

---

[8] Real exports, real imports and goods trade deficit are in million dollars. However, oil price is in dollars.

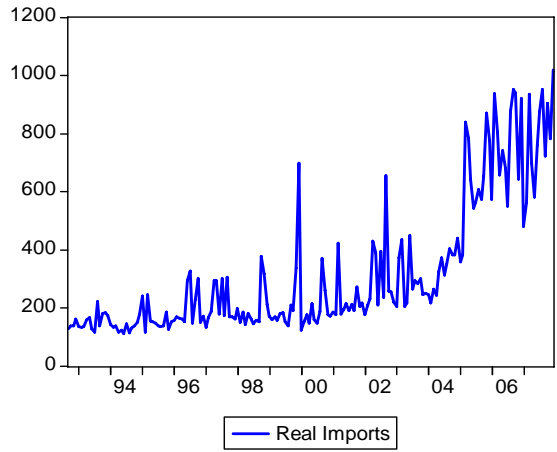
Real Imports

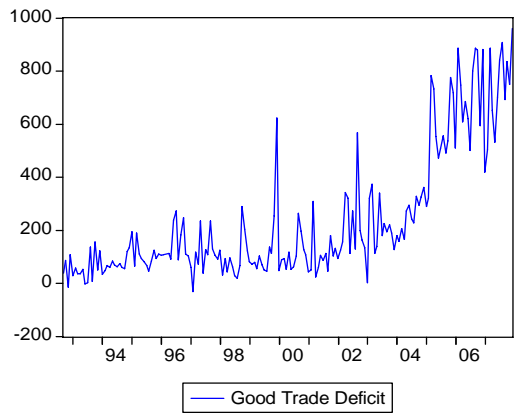
Good Trade Deficit

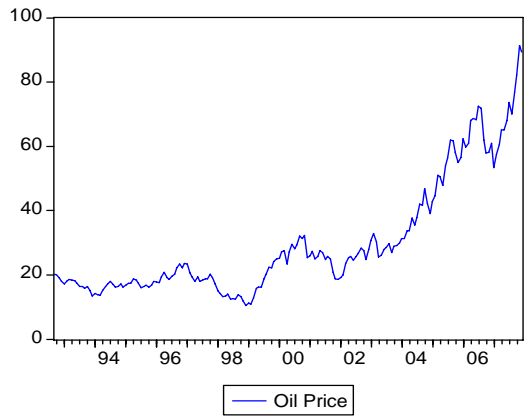
Oil Price

To test for unit root with structural breaks of real imports, goods trade deficit and oil price, we construct for each variable an equation with the same format as equation (1). The lag length of each equation is picked-up to guarantee the removal of the autocorrelation problem. Technically, we run for each equation two residual tests, which they are: Correlogram-Q-statistics and Serial Correlation LM tests. Both tests reject the hypothesis of autocorrelation for the three variables. Also, each equation is checked to assure the stability of the coefficients. Precisely, we use the cumulative sum of the least squares residuals (CUSUM) to check if we succeed to isolate the effect of the structural breaks on the parameters. All the estimations indicate we have stable parameters or stability in the equation over the sample period.[9]

The results of the unit root test are reported in Table (3). The main conclusion of the test is that real exports are stationary at the level. Additionally, real imports and goods trade deficit are stationary at the level if we take into consideration the problem of structural breaks. On the contrary, if we ignore the structural breaks, these two variables are stationary at the first difference. Meanwhile, oil price is stationary at the first difference only. It is useful to mention that the results illustrates that the structural breaks for the three variables are statistically significant. Hence, to inspect the potential impacts of oil price on real imports and real exports, we estimate a vector autoregressive model (VAR). This model includes oil price, real exports and real imports. Then, to capture the specific effect of oil price on goods trade deficit, we estimate a linear regression between the two variables.

**Table 3: ADF unit root test and unit root with structural breaks**

| Variable | ADF | | UR with Structural Break | |
|---|---|---|---|---|
| | Level | 1st diff. | Level | 1st diff. |
| Real Imports | -0.49 | -7.44*** | -10.35*** | - |
| Real Exports | -5.24*** | - | - | - |
| Goods Trade Deficit | -0.27 | -7.64*** | -9.83*** | - |
| Oil Price | 0.11 | -3.75** | -1.46 | -9.10*** |

Notes: 1) The ADF test is based on AIC.
2) */**/***: denotes significance at the 10/5/1 percent level, respectively.
3) The estimated lags of the ADF test are based on AIC.

**4. EMPIRICAL RESULTS**

**4.1 Impulse response functions**

We run the VAR model with three lags[10] by including the three variables in the model on the level. The results present the impact of oil price on the UAE real exports and real imports[11]. We employ the impulse response functions and the variance decomposition as the main tools to understand the influence of oil price on the UAE goods trade deficit with the U.S. Figure (4) presents the impulse response function of real exports and real imports to a unitary shock to the oil price for two years (24 months).

---

[9] The above-mentioned tests are not reported in the study, but they are available upon request.
[10] This number of lags guarantees no serial correlation, However, AIC advises two lags.
[11] To check the accuracy of the VAR model we run two residual tests of autocorrelation, which they are Portmanteau Autocorrelation and Serial LM Correlation LM Tests. Both tests cannot reject the null hypothesis of no residual autocorrelations up to lag 12. The results are attached in Appendix A.

### Figure 4: Dynamic effects of a unitary shock to oil price on real exports and real imports

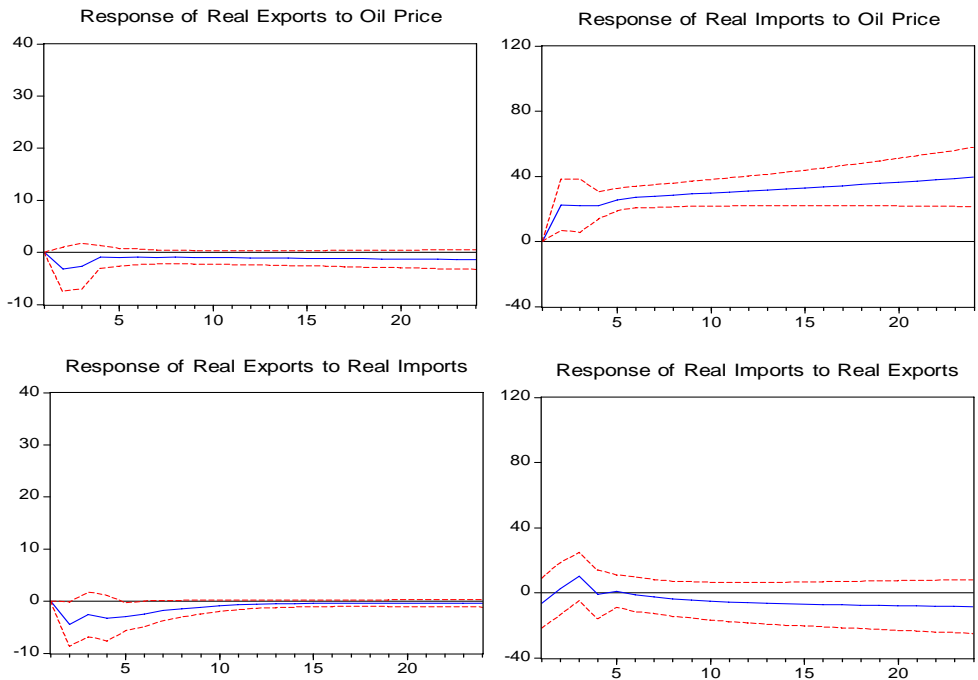

Response to Cholesky One S.D. Innovations ± 2 S.E.

The results illustrate that oil price has a lagged significant positive impact on imports at 1 percent significant level. However, the impact of oil price on real exports is statistically insignificant different from zero. Accordingly, the increase in oil price tends to expand the UAE goods trade deficit. To check-up this finding, we plot a *normalized* relationship between the two variables. Figure (5) illustrates an obvious positive relationship between oil price and goods trade deficit.

### Figure 5: The relationship between oil price and goods trade deficit

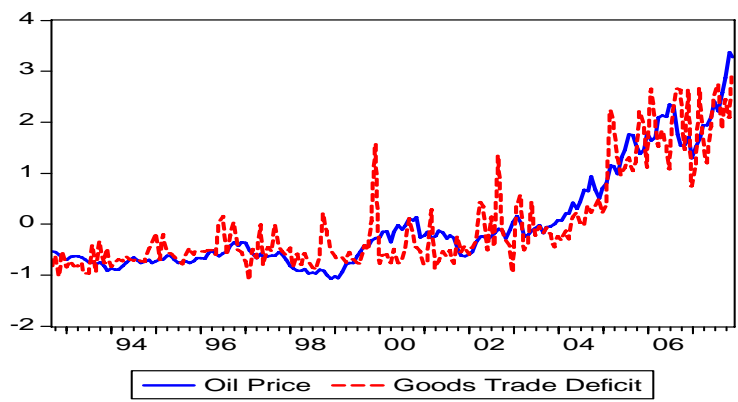

Then, in the second step we run an ARMA (1,1) regression model that includes oil price and goods trade deficit and has the following form:

$$GTD_t = \alpha + \sum_{i=0}^{k} \beta_i OP_{t-i} + \lambda GTD_{t-1} + \delta \varepsilon_{t-1} + \varepsilon_t \qquad (2)$$

Where $GTD_t$ is the goods trade deficit, $OP_t$ stands for the oil price, $\varepsilon_t$ is the stochastic errors. The lag of the ARMA (1,1) is determined based on the AIC, moreover, we run Correlogram-Q-statistics and Serial Correlation LM tests, both tests reject the hypothesis of autocorrelation12. The results are reported in Table (4) and confirm that oil price has a positive influence on goods trade deficit with a parameter equal to 4.8%. This effect is statistically significant different from zero and instantaneous.

**Table 4: The results of the ARMA (1,1) model**

| Parameter | Value | Standard Error | T- Value |
|---|---|---|---|
| $\alpha$ | -20.6 | 12.1 | -1.70 |
| $\beta_0$ | 4.8 | 2.3 | 2.06** |
| $\beta_1$ | -2.8 | 2.8 | -1.01 |
| $\lambda$ | 0.85 | 0.09 | 9.86*** |
| $\delta$ | -0.75 | 0.11 | -6.55*** |
| $R^2$-Adj | 0.81 | | |
| F-Stat. | 194.5 | | |

*/**/***: denotes significance at the 10/5/1 percent level, respectively.

One logical explanation for such behavior is that oil price is significant revenue or source of income for the UAE economy. Accordingly, the increase in the level of income stimulates more spending on imports. This attracts large number of American companies in the U.S. market to seek profitable investments and profitable contracts with the UAE economy. Due to the lack of time series data of the UAE economy and to support the abovementioned findings, the current study calculates using the available data the coefficient of variation[13] of oil price and real gross domestic by the type of expenditures. This includes the following variables: real private consumption, real government consumption and investment, real private investment, real imports and real exports[14]. We estimate the coefficient of variation for two different annual periods of time; (1993-1999) and (2000-2008). The second phase is the period of time at which price of oil starts to rise-up. The results are reported in Table (5) and show clearly that variation of the oil price creates significant variations in real private consumption, real private investment and real imports and a modest variation in real government consumption and investment. However, it has no impact on real exports.

---

[12] The results of those two tests are attached in Appendix B.
[13] It is the ratio of standard deviation over the mean.
[14] We exclude exports of oil and re-exports.

**Table 5: Coefficient of variation**

|           | Oil Price | RGDP | Private Cons. | Public Exp. | Private Investment | Imports | Exports |
|-----------|-----------|------|---------------|-------------|--------------------|---------|---------|
| 1993-1999 | 0.16      | 0.09 | 0.12          | 0.09        | 0.15               | 0.14    | 0.50    |
| 2000-2008 | 0.55      | 0.33 | 0.40          | 0.14        | 0.56               | 0.46    | 0.40    |

Another logical significant point is that even though oil revenue is a significant source to finance and to support the U.S. economy through buying U.S. treasury securities. Figure (6) shows the relationship between price of oil and the outstanding balance of the U.S. treasury securities for the oil exporting countries.[15] It is obvious that this relationship is strongly positive with a correlation coefficient equal to 0.93. The value of oil exporting countries securities increases from $45.2 billion in March 2000 to $137.9 billion in December 2007 to $213.9 billion in October 2010[16].

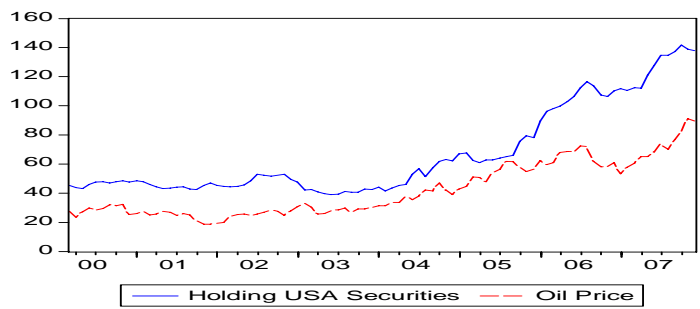

**Figure 6: The trend of oil price and the US securities hold by oil exporting countries**

The other key point of the results is to highlight on the relationship between real imports and real exports of the UAE. The primarily expectation of the UAE economy is to find a positive relationship between real imports and real exports. This anticipation is consistent with the vision of the UAE policymaker, which focuses on utilizing the oil revenues to support non-oil exporting sectors. Nevertheless, figure (7) presents real exports and real imports together. It is obvious that no clear relationship exists between them. However, we can tell that the two series are independent and would drift too far apart, particularly starting from 2003. Furthermore, figure (4) confirms that the dynamic response of real export for a shock in real import is negative. To the apposite, the dynamic effect of real imports for a shock in real exports fluctuates. The Impact of imports on exports is statistically significant but with a very small parameter. Furthermore, the variance decomposition in Table (5) confirms that real imports are unable to justify more than 3 percent of the fluctuations in real exports. At the same time, real exports are unable to explain more than 6 percent of the variations in real imports. Hence, the relationship between those two variables is very weak.

---

[15] These countries include Ecuador, Venezuela, Indonesia, Bahrain, Iran, Iraq, Kuwait, Oman, Qatar, Saudi Arabia, the United Arab Emirates, Algeria, Gabon, Libya, and Nigeria.

[16] Source of data is the US Department of the Treasury, available on line at http://www.treasury.gov/resource-center/data-chart-center/tic/Pages/ticsec2.aspx#ussecs.

**Figure 7: Real exports and real imports of the UAE**

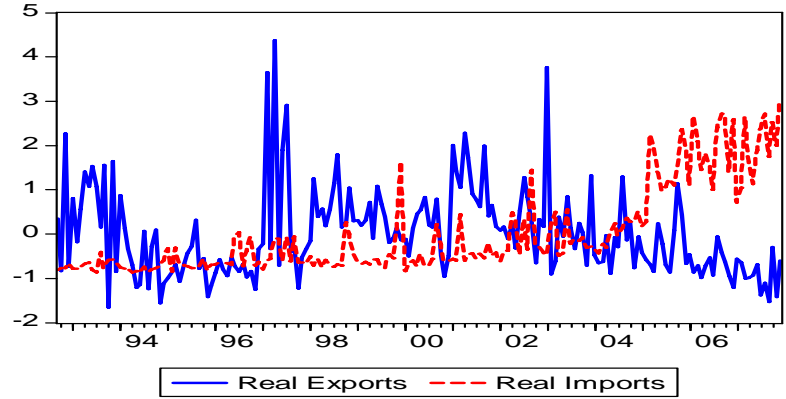

### 4.2 Variance Decompositions

The variance decomposition illustrates the portion of the forecast error variance of each variable in the model that can be explained by its own shocks and the exogenous shocks to the other variables in the model. Table (6) presents the variance decomposition of the VAR model. A major conclusion of the table is that oil price plays a significant role to explain the variation of real imports. The contribution of oil prices that explains real imports fluctuations is approximately 40.9 percent after one year and 62.8 percent after two years. This conclusion supports the above-mentioned finding, which is oil price of the UAE is a significant determinant of real imports. Empirically, it increases imports, as a result, goods trade deficit increases.

**Table 6: The variance decomposition of the VAR model**

| Dependent Variable | period | Export | Import |
|---|---|---|---|
| **Exports** | 1 | 100.0 | 0.0 |
|  | 12 | 91.5 | 5.9 |
|  | 24 | 89.7 | 6.0 |
| **Imports** | 1 | 0.4 | 99.6 |
|  | 12 | 1.5 | 57.7 |
|  | 24 | 2.7 | 34.5 |
| **Oil Price** | 1 | 0.0 | 0.0 |
|  | 12 | 2.5 | 40.8 |
|  | 24 | 4.2 | 62.8 |

## 5. CONCLUSIONS

The relationship between oil price and macroeconomic performance has attracted the attention of large numbers of economic scholars since the Arab oil embargo (1973-1974). During this period, economists succeeded to find different channels of how change in price of oil is transmitted to the economic activity.

The current paper seeks to comprehend the effect of oil price on the UAE goods trade deficit. The contribution of the current paper is to explore the impact of oil price shocks in a developing-oil-exporting country. The main motive behind the current paper is the necessity to understand the influence of oil price on the trade deficit. Beside, the UAE database lacks the required information to conduct such a significant research because it lacks long time series data. Hence, we claim, based on our knowledge, that this is the first study which investigates this issue in the UAE economy. As a result, the paper relies on data disseminated by the U.S. Census Bureau regarding trade in goods between the U.S. and the UAE. What's more, we utilize export and import price indexes from the Bureau of Labor Statistics to calculate real exports and real imports of the UAE economy. We know that our paper is focusing on around 11% of the UAE foreign trade. Nevertheless, we recognize that the U.S. is a major trade partner with the UAE.

The results of the current paper are interesting. We find a positive statistical significant relationship between oil price, real imports and goods trade deficit. The goods trade deficit increases as price of oil increases. Unexpectedly, oil price does not have a significant effect on real exports. The estimated coefficient of variation for real gross domestic product by type of expenditure supports strongly these conclusions. The findings of the current paper help to tell a story regarding price of oil and the performance of the UAE economy. There is no doubt that price of oil is a major determinant of the UAE economic activity. This outcome is consistent with the findings of Lorde et al. (2009) who explored the macroeconomic effect of oil price fluctuation in the same environment which is small open oil producing countries i.e. Trinidad and Tobago. The estimation of coefficient of variation provides an evidence that the revenue of oil is utilized to finance mainly private and government consumption which contributes to around 42% of gross domestic product during the period (1993-2008). At the same time, the increase in price of oil attracts more investment in the services sectors which contributes to approximately 12% of gross domestic product during the same period.

The policy implication of this finding is that the UAE economic policymakers do not devote direct portion of oil price revenues to encourage manufactured production and real exports. The encouragement could be indirectly through creating a promised investment environment and solid infrastructure. Furthermore, real exports and real imports are independent and they might drift far away from each other and cause larger goods trade deficit than the present figures, if the oil price keep going up-ward. This is a significant implication because it implies that the positive benefits from the increase in oil prices eliminated by increase in consumption and non-real productive spending.

**ACKNOWLEDGEMENTS**
The authors would like to acknowledge that this paper was funded by the University of Sharjah.

## *Appendix A*

**Table 7: The residual's test of the VAR model**

| VAR Residual Portmanteau Tests for Autocorrelations | | | | | |
|---|---|---|---|---|---|
| H0: no residual autocorrelations up to lag h | | | | | |
| Sample: 1992:09 2007:12 | | | | | |
| Included observations: 181 | | | | | |
| Lags | Q-Stat | Prob. | Adj Q-Stat | Prob. | df |
| 1 | 0.325069 | NA* | 0.326875 | NA* | NA* |
| 2 | 1.008903 | NA* | 1.018350 | NA* | NA* |
| 3 | 3.561894 | NA* | 3.614369 | NA* | NA* |
| 4 | 12.65423 | 0.1789 | 12.91218 | 0.1666 | 9 |
| 5 | 20.73211 | 0.2931 | 21.21954 | 0.2685 | 18 |
| 6 | 27.98829 | 0.4116 | 28.72451 | 0.3744 | 27 |
| 7 | 35.50968 | 0.4917 | 36.54848 | 0.4432 | 36 |
| 8 | 47.15339 | 0.3846 | 48.73063 | 0.3254 | 45 |
| 9 | 55.09270 | 0.4331 | 57.08537 | 0.3612 | 54 |
| 10 | 65.97219 | 0.3745 | 68.60109 | 0.2932 | 63 |
| 11 | 78.61111 | 0.2776 | 82.05782 | 0.1957 | 72 |
| 12 | 82.15311 | 0.4433 | 85.85133 | 0.3351 | 81 |
| *The test is valid only for lags larger than the VAR lag order. | | | | | |

**Table 8: The residual's test of the VAR model**

| VAR Residual Serial Correlation LM Tests | | |
|---|---|---|
| H0: no serial correlation at lag order h | | |
| Sample: 1992:09 2007:12 | | |
| Included observations: 181 | | |
| Lags | LM-Stat | Prob |
| 1 | 9.522461 | 0.3905 |
| 2 | 3.873487 | 0.9195 |
| 3 | 9.847107 | 0.3630 |
| 4 | 9.697412 | 0.3755 |
| 5 | 8.945010 | 0.4424 |
| 6 | 8.175392 | 0.5166 |
| 7 | 8.103747 | 0.5237 |
| 8 | 13.79097 | 0.1300 |
| 9 | 9.110187 | 0.4272 |
| 10 | 13.63197 | 0.1360 |
| 11 | 14.52701 | 0.1048 |
| 12 | 4.105069 | 0.9044 |
| Probs from chi-square with 9 df. | | |

## *Appendix B*

**Table 9: The residual's tests of the ARMA model**

| Correglogram-Q-Statistics Test | | | | |
|---|---|---|---|---|
| Sample: 1992:10 2007:12 | | | | |
| Included observations: 183 | | | | |
| Lags | AC | PAC | Q-Stat | Prob |
| 1 | 0.072 | 0.072 | 0.9542 | |
| 2 | -0.072 | -0.078 | 1.9323 | 0.165 |
| 3 | 0.006 | 0.017 | 1.9390 | 0.379 |
| 4 | -0.044 | -0.052 | 2.3021 | 0.512 |
| 5 | -0.047 | -0.038 | 2.7173 | 0.606 |
| 6 | 0.027 | 0.026 | 2.8518 | 0.723 |
| 7 | 0.010 | 0.000 | 2.8695 | 0.825 |
| 8 | -0.084 | -0.083 | 4.2354 | 0.752 |
| 9 | 0.060 | 0.071 | 4.9387 | 0.764 |
| 10 | 0.193 | 0.175 | 12.256 | 0.199 |
| 11 | 0.033 | 0.020 | 12.469 | 0.255 |
| 12 | -0.019 | -0.007 | 12.543 | 0.324 |
| 13 | -0.005 | -0.004 | 12.549 | 0.403 |
| 14 | -0.085 | -0.064 | 14.002 | 0.374 |
| 15 | -0.083 | -0.063 | 15.406 | 0.351 |
| 16 | -0.065 | -0.087 | 16.273 | 0.364 |
| 17 | -0.015 | -0.011 | 16.319 | 0.431 |
| 18 | 0.074 | 0.094 | 17.450 | 0.424 |
| 19 | -0.039 | -0.082 | 17.760 | 0.472 |
| 20 | -0.072 | -0.107 | 18.835 | 0.467 |
| 21 | -0.079 | -0.093 | 20.128 | 0.450 |
| 22 | -0.159 | -0.171 | 25.417 | 0.230 |
| 23 | -0.086 | -0.092 | 26.970 | 0.212 |
| 24 | 0.060 | 0.060 | 27.738 | 0.226 |
| 25 | -0.091 | -0.091 | 29.528 | 0.201 |
| 26 | -0.128 | -0.093 | 33.071 | 0.129 |
| 27 | -0.012 | -0.048 | 33.100 | 0.159 |
| 28 | 0.039 | -0.023 | 33.431 | 0.183 |
| 29 | 0.083 | 0.091 | 34.963 | 0.171 |
| 30 | 0.015 | -0.001 | 35.016 | 0.204 |
| 31 | 0.019 | 0.057 | 35.097 | 0.239 |
| 32 | -0.024 | 0.086 | 35.228 | 0.275 |
| 33 | 0.107 | 0.171 | 37.808 | 0.221 |
| 34 | -0.048 | -0.095 | 38.339 | 0.240 |
| 35 | 0.012 | 0.063 | 38.371 | 0.278 |
| 36 | 0.065 | 0.078 | 39.331 | 0.282 |

**Table 10: The residual's test of the ARMA model**

| Breusch-Godfrey Serial Correlation LM Test: | | | |
|---|---|---|---|
| F-statistic | 0.734861 | Probability | 0.660590 |
| Obs*R-squared | 6.116757 | Probability | 0.634155 |
| | | | |
| Test Equation: | | | |
| Dependent Variable: RESID | | | |
| Method: Least Squares | | | |
| Presample missing value lagged residuals set to zero. | | | |
| Variable | Coefficient | Std. Error | t-Statistic | Prob. |
| C | 4.351341 | 13.08133 | 0.332638 | 0.7398 |
| OP | 0.848919 | 2.509899 | 0.338228 | 0.7356 |
| OP(-1) | -1.257204 | 3.142365 | -0.400082 | 0.6896 |
| DEF(-1) | 0.031631 | 0.094916 | 0.333257 | 0.7394 |
| MA(1) | 0.511018 | 0.486914 | 1.049503 | 0.2954 |
| RESID(-1) | -0.463264 | 0.471441 | -0.982655 | 0.3272 |
| RESID(-2) | -0.495026 | 0.355344 | -1.393090 | 0.1654 |
| RESID(-3) | -0.300437 | 0.270841 | -1.109273 | 0.2689 |
| RESID(-4) | -0.290693 | 0.207598 | -1.400269 | 0.1633 |
| RESID(-5) | -0.210625 | 0.161658 | -1.302911 | 0.1944 |
| RESID(-6) | -0.115828 | 0.133568 | -0.867187 | 0.3871 |
| RESID(-7) | -0.085364 | 0.109558 | -0.779174 | 0.4370 |
| RESID(-8) | -0.165602 | 0.099708 | -1.660878 | 0.0986 |
| R-squared | 0.033425 | Mean dependent var | | -0.098606 |
| Adjusted R-squared | -0.034804 | S.D. dependent var | | 105.3898 |
| S.E. of regression | 107.2081 | Akaike info criterion | | 12.25581 |
| Sum squared resid | 1953910. | Schwarz criterion | | 12.48381 |
| Log likelihood | -1108.407 | F-statistic | | 0.489894 |
| Durbin-Watson stat | 1.996583 | Prob(F-statistic) | | 0.918694 |